\documentclass[aps,pre,twocolumn,showpacs,superscriptaddress,amsmath,amssymb]{revtex4}

\usepackage{graphicx}
\usepackage{multirow}

\begin{document}

\title{Time evolution of predictability of epidemics on networks}

\author{Petter Holme}
\email{holme@skku.edu}
\affiliation{Department of Energy Science, Sungkyunkwan University, Suwon 440--746, Korea}
\affiliation{Department of Physics, Ume{\aa} University, 90187 Ume\aa, Sweden}
\affiliation{Department of Sociology, Stockholm University, 10961 Stockholm, Sweden}

\author{Taro Takaguchi}
\affiliation{National Institute of Informatics, 2-1-2 Hitotsubashi, Chiyoda-ku, Tokyo, 101-8430, Japan}
\affiliation{JST, ERATO, Kawarabayashi Large Graph Project, 2-1-2 Hitotsubashi, Chiyoda-ku, Tokyo, 101-8430, Japan}

\begin{abstract}
Epidemic outbreaks of new pathogens, or known pathogens in new populations, cause a great deal of fear because they are hard to predict. For theoretical models of disease spreading, on the other hand, quantities characterizing the outbreak converge to deterministic functions of time. Our goal in this paper is to shed some light on this apparent discrepancy. We measure the diversity of (and, thus, the predictability of) outbreak sizes and extinction times as functions of time given different scenarios of the amount of information available. Under the assumption of perfect information---i.e.,\ knowing the state of each individual with respect to the disease---the predictability decreases exponentially, or faster, with time. The decay is slowest for intermediate values of the per-contact transmission probability. 
With a weaker assumption on the information available, assuming that we know only the fraction of currently infectious, recovered, or susceptible individuals, the predictability also decreases exponentially most of the time. There are, however, some peculiar regions in this scenario where the predictability decreases. In other words,  to predict its final size with a given accuracy, we would need increasingly more information about the outbreak.
\end{abstract}


\maketitle

\section{Introduction}

Outbreaks of serious infectious diseases can be very frightening, even though many times the diseases die out before reaching the people who are worryed about it. Part of the reason for this is that it is hard to forecast disease outbreaks~(see e.g.\ Refs.~\cite{hard_to_predict1,hard_to_predict2}), and this uncertainty adds to the fear. At the same time, our common models for disease spreading behave, once they take hold in the population~\cite{unpredictability},  like deterministic, perfectly predictable, quantities~\cite{janson_etal}. Such models have two components. First, they model the person-to-person contagion and the history of one individual with respect to the disease. This is done by dividing individuals into states (or ``compartments'') with respect to the disease, and assigning rules for transitions between the states. A canonical model of diseases that make the infectious person immune upon recovery (or kill their host) is the Susceptible-Infectious-Recovered (SIR) model. This model has three classes: susceptible individuals can acquire the disease, infectious individuals can spread it further, and recovered individuals (who, technically speaking, can also be dead) cannot get the disease nor can they spread it. The transition rules are that a susceptible individual, upon meeting an infectious individual, can (with some probability) become infectious, and after some time, or with some probability, that individual recovers or dies. These transitions are assumed to be instantaneous (which is quite a coarse simplification with respect to real diseases). The second component of epidemic models describes the population level processes over which the pathogens propagate. For decades, theoretical epidemiologists have ignored this issue and taken the ``well-mixed assumption,'' i.e., that any two individuals have the same chance of meeting during an interval of time. Recently, researchers have recognized this problem and represented the contact structure as a network~\cite{nwkepi_keeling,nwkepi_morris,nwkepi_pastor}. The regularities, or \emph{structure}, of the network have a great impact on the evolution of disease outbreaks. Among the network structures, the one with the strongest influence is perhaps the degree distribution---the probability distribution of the number of neighbors. This structure explained the existence of super-spreaders~\cite{superspreader} and challenged the existence of an epidemic threshold below which an outbreak would always die out quickly~\cite{no_threshold}. Now,  if one studies the SIR process on the configuration model~\cite{Molloy}---random networks with an arbitrary degree distribution---the outbreak turns out to be completely predictable in the long-time and large-$N$ limits~\cite{janson_etal}. To be specific, population-averaged quantities such as the fraction of infected individuals converge to deterministic functions of time. The major uncertainty is in the very beginning, whether the outbreak would die out immediately or not. There is thus an apparent contradiction---our canonical compartmental models seem unable to capture the uncertainty of real-world outbreaks.

In this work, we investigate how the predictability of SIR processes evolves with the outbreak itself. We study questions such as the following: Assuming an outbreak started some time $t$ ago, then what can we say about the final number of infected people $\Omega$ and the extinction time $\tau$ as a function of $t$? How does our ability to predict $\Omega$ and $\tau$ depend on the network structure and the type of information we have about the outbreak?
In particular, we consider the two senarios in which the state of each individual is known (see Fig.~\ref{fig:ill} for a schematic illustration and Sec.~\ref{sec:pred_state}) and only the fractions of currently infectious, recovered, or susceptible individuals are known (see Sec.~\ref{sec:pred_comp}).

\section{Preliminaries}

In this section, we will clarify the methods and precise model definitions used in the rest of the paper. We also mention some computational considerations.

\subsection{SIR simulation}

We assume there is a disease spreading over a static underlying network represented as a graph, $G=(V,E)$. Set $V$ is a set of vertices representing individuals of the population; $E$ is a set of edges representing (unordered) pairs of individuals in close enough contact for the disease to spread. The number of vertices is denoted by $N$ and the number of edges is by $M$. The vertices are, at a given time, in one of three states---S (meaning that they do not have the disease, but they can get it), I (meaning that they have the disease and they can spread it), and R (meaning that they do not have the disease and they cannot get it). We assume a disease outbreak starts at $t=0$. At the beginning, all vertices belong to the state S, except for a randomly chosen vertex that is in state I. If a pair $(i,j)\in E$ consists of one I and one S vertex, then the S vertex has a chance $\beta$ of becoming I (otherwise it stays S).  We call this an \emph{infection event}.  Every vertex with state I has a chance $\nu$ per time unit of becoming R (in a \emph{recovery event}). The state of the system at a certain time could thus be fully described by two vectors: ${\bf s}=(s_1,\dots,s_N)$ giving the state of each vertex ($s_i\in\{\mathrm{S},\mathrm{I},\mathrm{R}\}$), and a vector $\mathbf{t}_\mathrm{I}=(t^\mathrm{I}_1,\dots,t^\mathrm{I}_N)$ giving the infection time of the vertices.

Most important quantities describing the outbreak will only depend on the ratio between $\beta$ and $\nu$ (in the well-mixed SIR model, this ratio is called $R_0$, but to not confuse things, we do not use this name). The actual values of $\beta$ and $\nu$  are only needed to calculate the real time to reach the peak prevalence, extinction, etc. That does not interest us in the present paper, so we measure time in units of $1/\nu$. In a simulation, this can conveniently be done by (at an iteration of the algorithm) performing a random infection event with a probability
\begin{equation}\label{eq:prob}
P= \frac{\beta\Sigma}{\beta\Sigma + I},
\end{equation}
where $\Sigma$ is the number of edges between an I and an S vertex, and $I$ is the prevalence (note that roman letters symbolize a state and italicized letters represent the number of vertices in the corresponding state)~\cite{holme_versions}. The factor $\Delta t = (\beta \Sigma + I)^{-1}$ represents the time increment since the last iteration. Thus, to keep track of the time (in units of $1/\nu$), one adds $\Delta t$ to a variable representing time. If an infection event is not performed (which happens with a probability $1-P$), we perform a recovery event. In an infection event, the SI edge (to become II) is chosen randomly among all SI edges. Similarly, in a recovery event, the I vertex (to become R) is selected with uniform randomness among all I vertices.

\begin{figure*}
\includegraphics[width=0.9\textwidth]{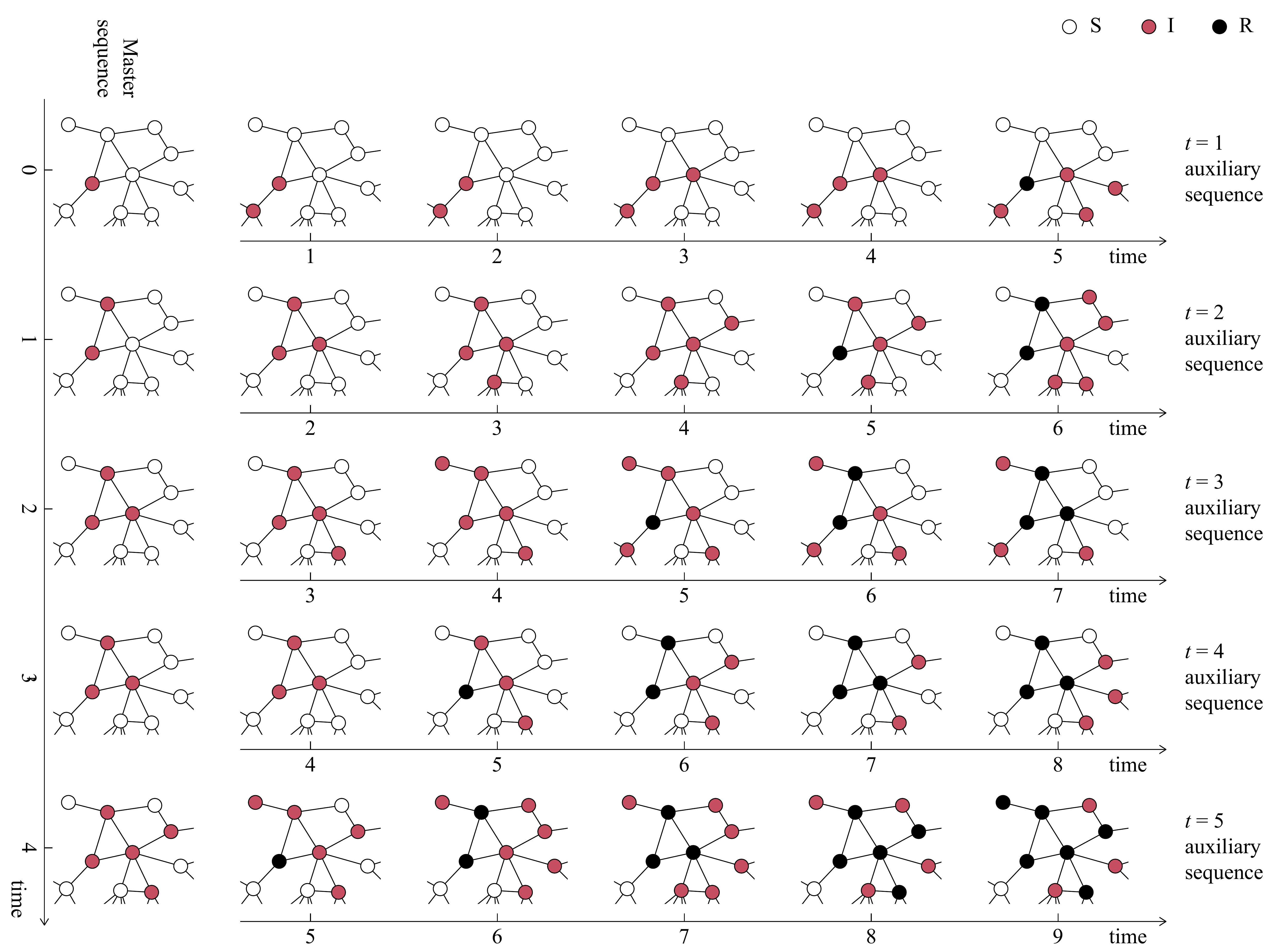}
\caption{(Color online) Illustration of the methodology to study the decay of predictability in the case of maximal information of the system. First we fun a master sequence simulation of the SIR process on a network. At every time step, we break this simulation and use the configuration as a seed for $1{,}000$ auxiliary simulations. The standard deviation of the final outbreak size and time to extinction thus captures how accurately the outbreak can be predicted given the state of the system at the breaking point.
}
\label{fig:ill}
\end{figure*}

\begin{figure}
\includegraphics[width=0.7\columnwidth]{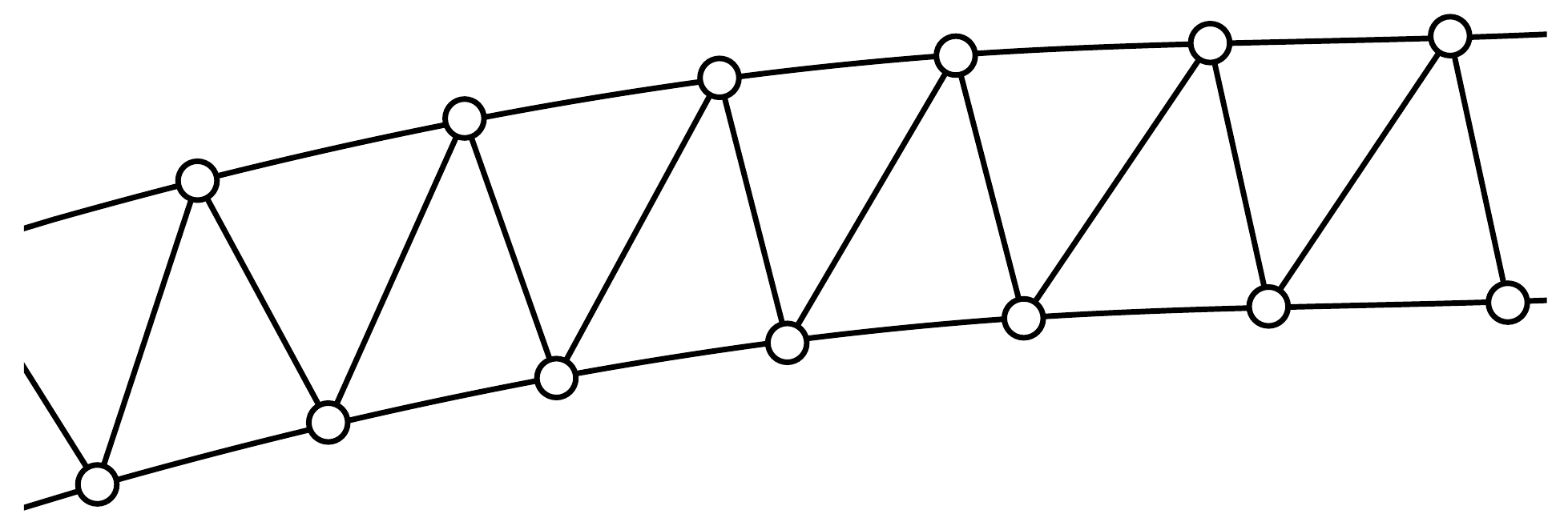}
\caption{The local structure of the Watts-Strogatz model networks with $k=4$.
}
\label{fig:sw_ill}
\end{figure}

\subsection{Network models}

To study the effects of network topology on the outbreak predictability, we use a collection of six network models. We choose the models to reflect a variety of stylized network topologies. This methodology is inspired by Ref.~\cite{topocollection}. Throughout this paper, we will use graphs of size $N=2{,}500$. We use $1{,}000$ realizations of each network model. A summary of the basic network structure in these networks can be found in Table~\ref{tab:stats}.

\subsubsection{Large- and small-world networks}

First, we use the Watts-Strogatz small-world network model~\cite{ws}. In this model, all vertices have the same degree $k$ (number of neighbors). They are initially arranged on a ring and connected to their $k/2$ nearest neighbors, at either side, of the ring. Counting modulo $N$, these neighbors can be enumerated as $i-k/2,i-k/2+1,\dots,i-1,i+1,\dots,i+k/2$.  Then one goes through all vertices $i$ around the ring, and for every edge $(i,j)$ pointing to a neighbor ahead (i.e.,\ $j>i$ modulo $N$), with probability $p$, one replaces (or ``rewires'') it by $(i,j')$ (where $j'$ is a different vertex from $i$ and $(i,j')\notin E$). By construction, if $p=0$ and $k>2$, this model gives networks with a high density of triangles. Furthermore, for these parameter values, the average distances grows like $N$. Watts and Strogatz show that if  $p$ is slightly larger than zero, the distances scales much slower with $N$. The number of triangles is, on the other hand, not so sensitive to $p$. The reason for this behavior is that the rewired edges (usually called ``long-range edges,'' with reference to the ring topology) connect distant parts of the graph. For the triangles, only the directly involved vertices are affected by a rewiring, but for the distances the extended neighborhoods of these vertices are affected. In this work, we use the Watts-Strogatz model with $p=0$ (calling it \emph{large-world networks}) and $p=0.01$ (calling it \emph{small-world networks}). We use $M=5{,}000$, so the small-world networks have an expected number of 50 rewired edges. The large-world networks, of course, are all isomorphic and only one copy of them is needed as a substrate for our simulations.

\subsubsection{Random regular graphs}

Random regular graphs~\cite{rr} are designed to be as uniform as possible with respect to the positions of vertices. Like the Watts-Strogatz model, all vertices have the same degree $k=4$ (giving $M=5{,}000$). Other than that, they are as random as possible. In graph models with fewer restrictions (such as Erd\H{o}s-R\'enyi random graphs~\cite{mejn:book}, where there is a fixed probability for any vertex pair to belong to $E$), the degrees vary, which differentiates the vertices. On the other hand, for our purposes, Erd\H{o}s-R\'enyi graphs would probably give very similar results.

\subsubsection{Scale-free networks}

So far, all the model networks we discussed have uniform degree distributions. This is not very realistic. Rather, the degrees of empirical networks are in general often broadly and skewedly distributed. This is also true for some of the particular networks that diseases spread over, such as sexual networks~\cite{sexnetrev} or networks of contacts between patients in hospitals~\cite{hospital}. To model such networks, we use the configuration model~\cite{mejn:book} with an emerging power-law degree distribution. We draw $N$ integers $\{k_i\}_{i=1}^N$ from a probability distribution
\begin{equation}\label{eq:power_law}
P(k)=\left\{\begin{array}{ll} k^{-\gamma} & \text{if $1\leq k\leq N-1$,}\\
0 & \text{otherwise,} \end{array}\right. 
\end{equation}
until $\sum_{i=1}^Nk_i$ is even.
Then we randomly attach edges between vertices until vertex $i$ has $k_i$ neighbors. We do not forbid multiple edges or self-edges. That is, the resulting object is a multigraph. The upper limit $N-1$ is somewhat arbitrarily chosen to be the same as the maximum degree in a simple graph. This is needed to keep the fluctuations down (that comes from the extreme variation if $k$ is generated by small $\gamma$ values). Finally, we construct a simple graph by removing multiple edges and self-edges. We try three different $\gamma$ values from the typical range of empirical networks~\cite{caldabook}: $2$, $2.5$, and $3$.

\subsubsection{Theoretical threshold values}

As mentioned, an interesting feature of epidemic models is that they can have a threshold behavior where the average fraction of infected individuals is finite as $N\rightarrow\infty$ provided that $\beta$ is larger than a threshold value $\beta_c$. By analogy to models in statistical physics, this could be described as a continuous phase transition~\cite{nwkepi_pastor}. This is not the focus of our study, but we will review a few theoretical results for our models to give some context.

The large-world and scale-free networks have  thresholds for the extreme values of $\beta$. In the large-world network case, the epidemics spread along a one-dimensional chain (Fig.~\ref{fig:sw_ill}). For any $\beta<1$ value there is a finite chance that the outbreak will stop, so there is an expected distance it will propagate from the seed. The outbreaks are limited by this finite distance, so  if $N\rightarrow\infty$, the fraction of affected vertices is zero. For the scale-free networks of the exponents that we study, the threshold is $\beta_c=0$~\cite{no_threshold,nwkepi_pastor}. The random regular networks have a threshold $\beta_c=1/2$~\cite{nwkepi_pastor}. The small-world networks in the limit $p=1$ have $\beta_c=2/7$~\cite{yamir_etal}, but we are not aware of any derivations of the threshold for  $0<p<1$.

\begin{table}
\caption{\label{tab:stats} 
Summarizing the network structure of the six classes of networks we study. All networks have $2{,}500$ vertices. Except the large-world network (which is unique), we use the same $1{,}000$ independent realizations as in the rest of the work. In the second row we list the standard deviation (once again with the exception for the large-world network).}
\begin{ruledtabular}
\squeezetable
  \begin{tabular}{l|llll}
  network model & $M$ & $s$ & $d$ & $C$ \\\colrule
  large world & 5{,}000 & 1 & 312.8 & 0.5 \\ \colrule
  \multirow{2}{*}{small world} & 5{,}000 & 1 & 36.6  & 0.484 \\ & 0 & 0 & 3.69 & 0.00236  \\\colrule
  \multirow{2}{*}{random regular} & 5{,}000 & 1  &  6.47 &  0.000964 \\  & 0 & 0 & 0.00401  & 0.000440 \\\colrule
  \multirow{2}{*}{scale free, $\gamma=2$} & 6696 & 0.963 & 3.24&0.0666  \\ & 365 & 0.00783 & 0.125 &  0.0132\\\colrule
  \multirow{2}{*}{scale free, $\gamma=2.5$} & 3039 & 0.785 & 4.54 & 0.0153 \\ & 160 & 0.0271  & 0.464 & 0.00497\\\colrule
  \multirow{2}{*}{scale free, $\gamma=3$} & 2046 & 0.46 & 6.88 & 0.00277 \\ & 69.3 &  0.0459 &1.01 & 0.00161\\
  \end{tabular}
\end{ruledtabular}
\end{table}

\section{Predictability given the state of the system}\label{sec:pred_state}

The predictability of any kind of phenomenon depends on the information available and the ability to use it. In this paper, we assume that the disease spreading is determined by the SIR process. In addition, we assume that we know the precise state of the system, i.e., the underlying network, the state of all vertices, and when they changed state. In that case, one cannot, by definition, do better in predicting the future than evaluating the SIR process with the state of the system as the input. In this section, we focus on this limit of maximum information about an outbreak.

\subsection{Methods}

We want to understand how the predictability of $\Omega$ and $\tau$ depends on $t$ given that we know $\mathbf{s}$ and $\mathbf{t}^\mathrm{I}$. To this end, we run an SIR simulation producing a \textit{master sequence} $(\mathbf{s}_0,\mathbf{t}^\mathrm{I}_0)$, and at every time step, we start $1{,}000$ simulation runs with $(\mathbf{s}_0,\mathbf{t}^\mathrm{I}_0)$ as the initial condition. The standard deviations of  $\Omega$ and $\tau$ of these \textit{auxiliary sequences} measures the unpredictability evolution of the master sequence. Finally, we average these standard deviations over at least $2{,}000$ master sequences and call them $\sigma_\Omega$ (for the outbreak size) and $\sigma_\tau$ (for the extinction time). For each such sequence, we take a random network realization from a pool of $1{,}000$. When an outbreak is dead (when $\Sigma=0$), it contributes with zero to the average and standard deviation. This procedure is illustrated in Fig.~\ref{fig:ill}.

\begin{figure*}
\includegraphics[width=\textwidth]{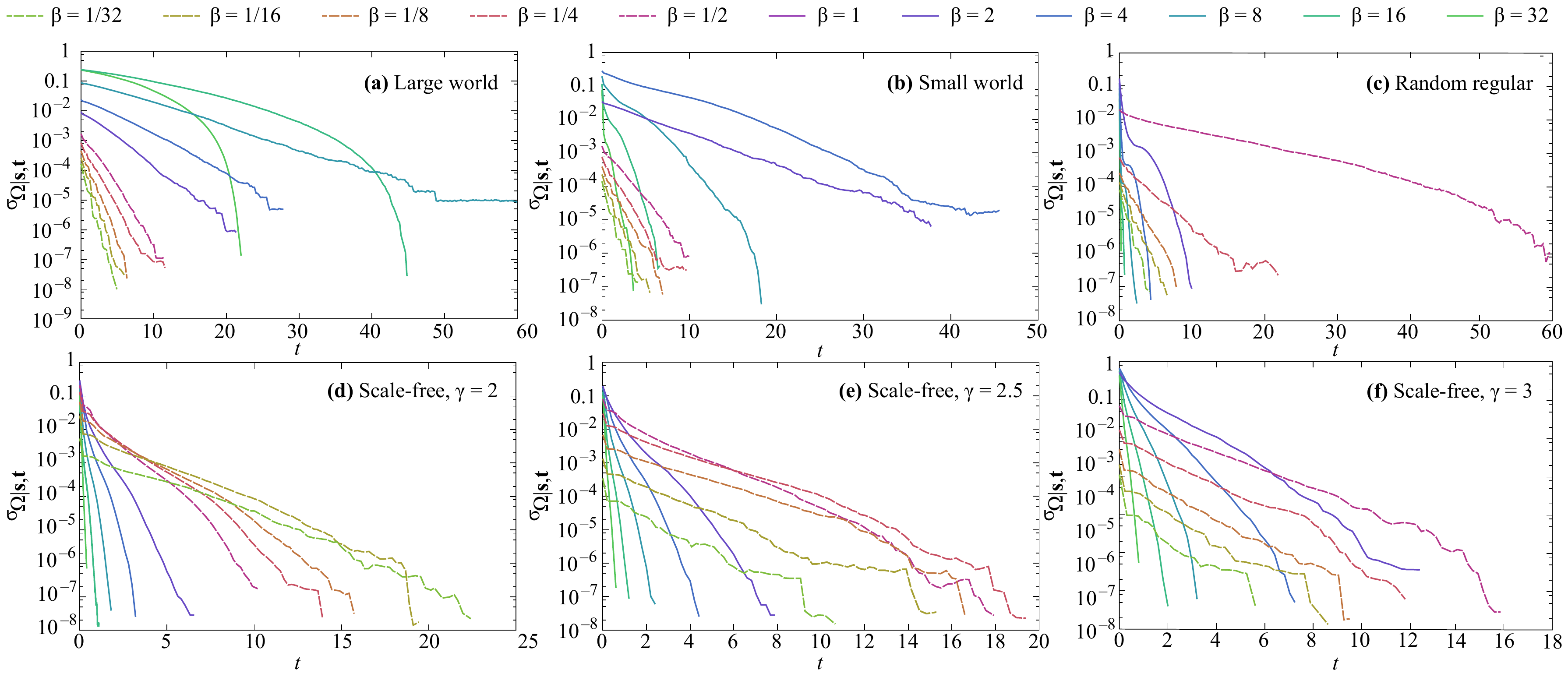}
\caption{(Color online) Standard deviation of the final outbreak size given the state at $t$, as a function of $t$. As $\sigma_\Omega$ measures the outbreak diversity, its increase reflects the decay of predictability. The curves ends when the all the simulated outbreaks have died. In principle $\sigma_\Omega$ is defined for any $t>0$, but after the outbreaks are over it is zero and thus not visible in this figure.
}
\label{fig:size}
\end{figure*}

\begin{figure*}
\includegraphics[width=\textwidth]{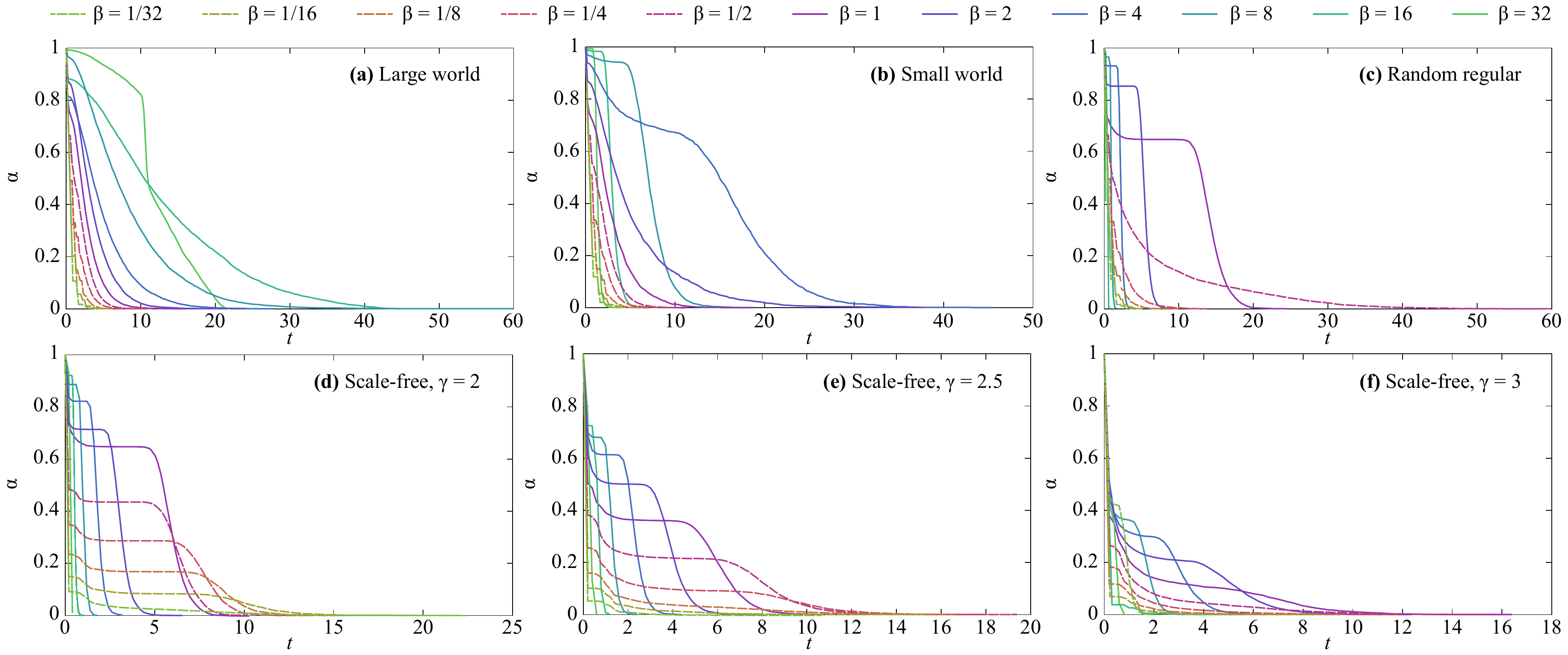}
\caption{(Color online) The fraction of surviving outbreaks as a function of $t$. The panels and parameter values are the same as in Fig.~\ref{fig:size}.
}
\label{fig:alive}
\end{figure*}

\begin{figure*}
\includegraphics[width=\textwidth]{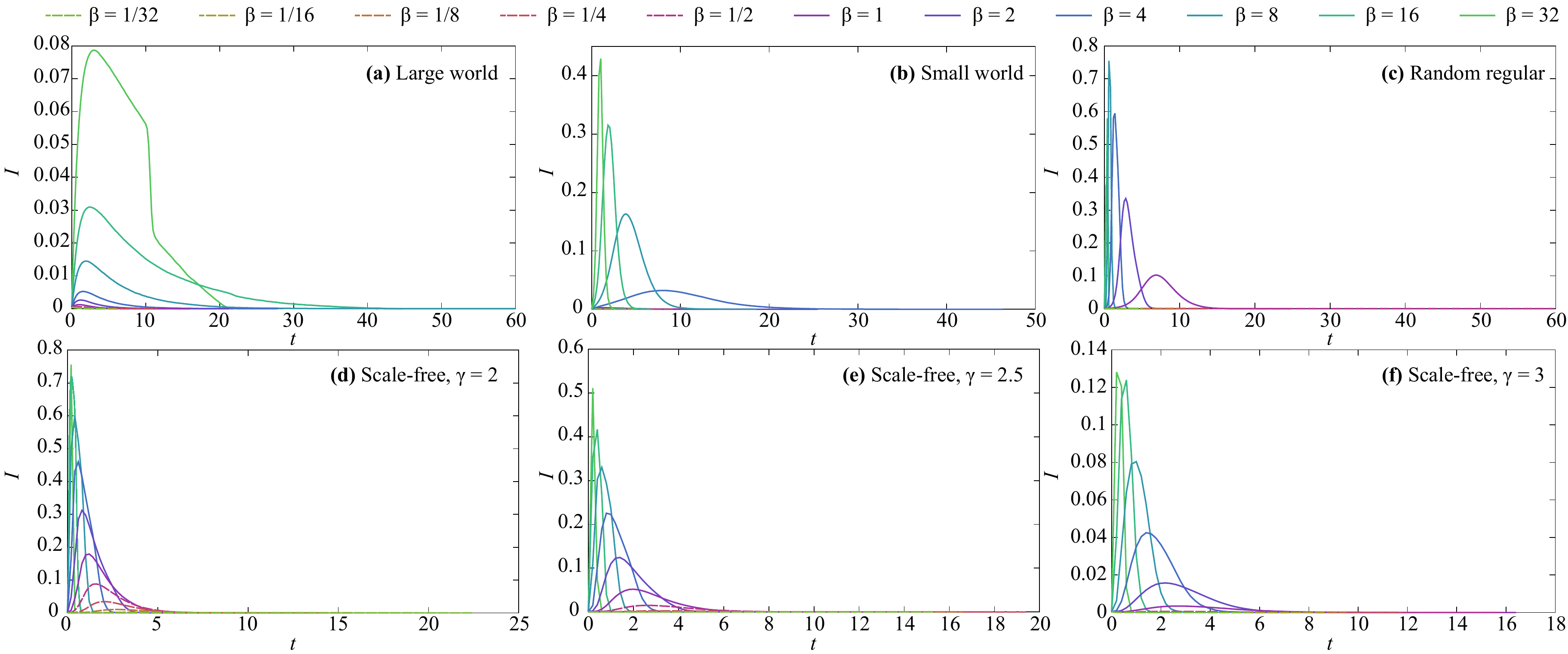}
\caption{(Color online) The prevalence (fraction of infectious vertices) as a function of $t$. The panels and parameter values are the same as in Fig.~\ref{fig:size}.
}
\label{fig:prev}
\end{figure*}

\begin{figure}
\includegraphics[width=\columnwidth]{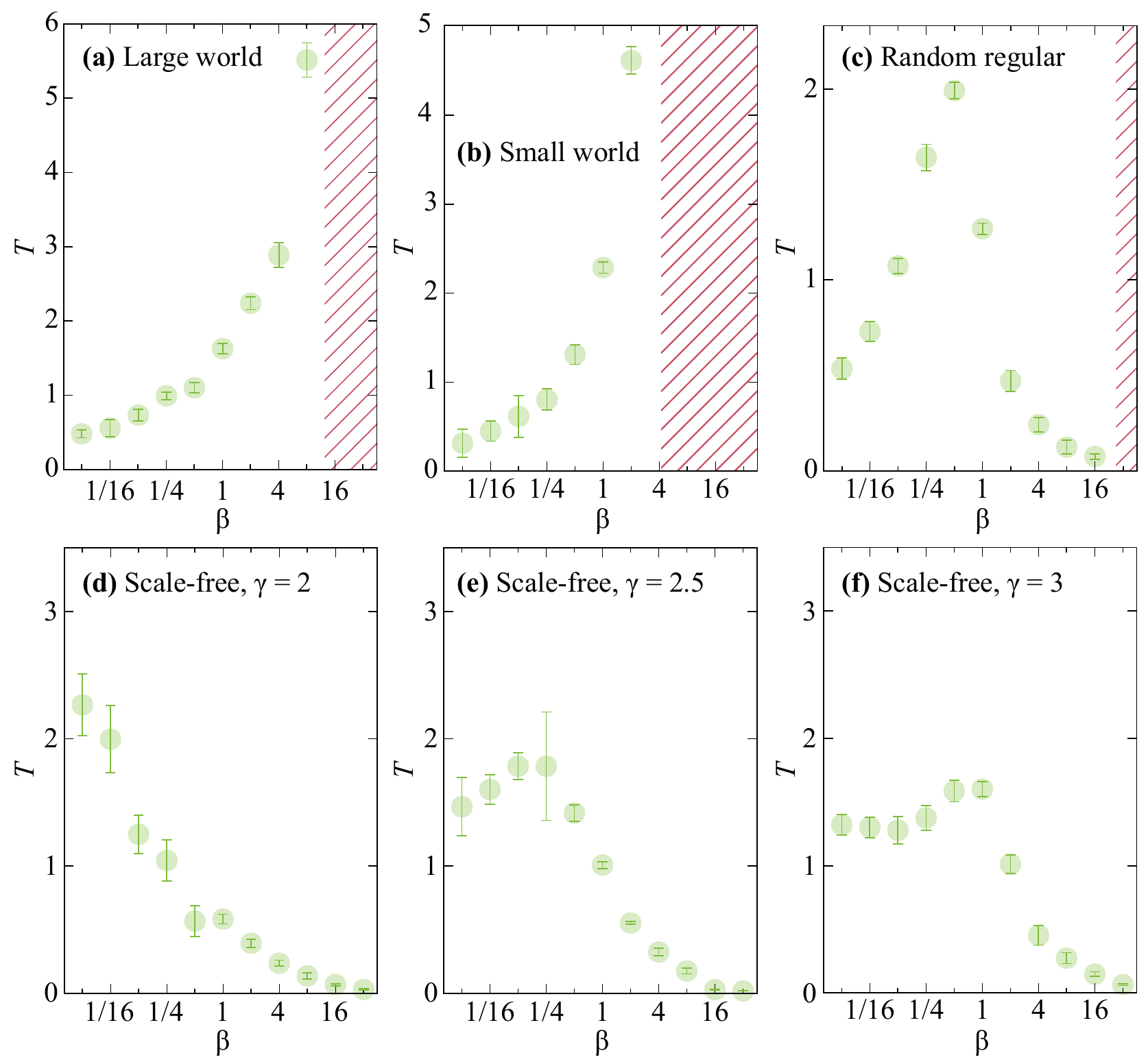}
\caption{(Color online) The decay constants from nonlinear least-squares fits to exponential forms of the curves shown in Fig.~\ref{fig:size}. The shaded regions indicate where $\sigma_\Omega$ does not fit an exponential function. The error bars represent standard errors.
}
\label{fig:t}
\end{figure}

\subsection{Results}

Now we will turn to our numerical results relating to the predictability.

\subsubsection{Predicting the final outbreak size}

We start by investigating the $\sigma_\Omega$ for our six classes of networks for an exponentially increasing sequence of 11 $\beta$ values. We choose the $\beta$ values so that we cover outbreaks of all sizes, in all network models. We plot the results in Fig.~\ref{fig:size}. Our first observation is that none of the curves (i.e.\ for no $\beta$ value and for no topology) decays slower than exponential. Most of them decay roughly exponentially, while some decay more rapidly.  The faster-than-exponential decay is clearest for the highest $\beta$ curves of the small- and large-world networks (see Figs.~\ref{fig:size}(a) and (b)). The random-regular graphs in Fig.~\ref{fig:size}(c) do not have the same fast drop-off. Since all three models in Figs.~\ref{fig:size}(a)--(c) have uniform degree distributions, the explanation must be something else. Locally, the large- and small-world networks look the same---bands of stacked triangles (see Fig.~\ref{fig:sw_ill}). We believe the cut-off of the exponential decay relates to the typical length of these bands. Thus, for example, the $\beta=8$ curve in Fig.~\ref{fig:size}(b) bends down around $t=18$. Our interpretation is that the early, slower decay is a period in which none of the outbreaks has reached the entire population. For the other models, since the outbreak can reach the entire population much faster, these two regions get blurred and the result is just one exponential decay. Comparing curves of different $\beta$ values, we see that the ones with the slowest decay are the ones with the longest extinction times. Presumably, as an extinct outbreak is perfectly predictable, the extinction time is an important factor in determining the time evolution of predictability. As a reference, to see the progression of the outbreak, we plot the fraction of surviving outbreaks as a function of time in Fig.~\ref{fig:alive} and the average prevalence in Fig.~\ref{fig:prev}. 
The increasing effect of perfectly predictable dead outbreaks would explain why intermediate $\beta$ values have the slowest decay of predictability---extinction times will initially grow with $\beta$ (as the chance for an early extinction decreases) and then decrease (due to the increasingly fast spread and subsequent burn-out in the population~\cite{extinction_time}). This is, however, not the case for the scale-free networks in the same parameter range. In Figs.~\ref{fig:size}(d)--(f), we plot the results for scale-free networks. For $\gamma = 2.5$ and $3$, when $\beta$ is small, the decay rate of $\sigma_\Omega$ is almost constant. The $\sigma_\Omega$ value still increases with $\beta$, but this increase happens in the early die-off (mentioned in the Introduction). In other words, smaller $\beta$ affects $\sigma_\Omega$ by increasing the chance that the disease will die in the very early stage. Once the outbreak takes hold in the population, the predictability decays independent of $\beta$. This can also be seen in Fig.~\ref{fig:alive} as plateaus after the initial die-off. For larger $\beta$, over the peak in extinction time, the lifetime decreases with $\beta$. One interesting thing to note from Fig.~\ref{fig:prev} is how long the outbreaks live after the peak. The prevalence for many of the curves in Fig.~\ref{fig:prev} has reached very low values before the corresponding curve in the fraction of surviving outbreaks (Fig.~\ref{fig:alive}).

To investigate more closely the relationship between the decay of $\sigma_\Omega$ and $\beta$, we assume the scaling form  $\sigma_\Omega\sim\exp(-t/T)$ and measure $T$~\cite{footnote}. Note that this assumption is not justified by any rigorous theory---it should be regarded as a somewhat sketchy summary of Fig.~\ref{fig:size}. In Fig.~\ref{fig:t}, we show the results. Some of the curves (as mentioned) do not have an exponential tail and are excluded from this analysis. This figure illustrates how, except for maybe the $\gamma=2$ scale-free network, all curves of Fig.~\ref{fig:size}  have the slowest decay of $\sigma_\Omega$ for intermediate $\beta$ values. It seems reasonable that the location of this peak converges to the epidemic threshold as $N\rightarrow\infty$. On the other hand, none of the Fig.~\ref{fig:size} curves has a concave shape indicative of a slower-than exponential decay (which one could expect if $\sigma_\Omega$ was a critical parameter, diverging at the threshold). An alternative hypothesis is that this peak coincides with the maximal extinction time, which is thought to be larger than, but distinct from, the epidemic threshold~\cite{extinction_time}. We will leave this as a question for future studies.

\begin{figure*}
\includegraphics[width=\textwidth]{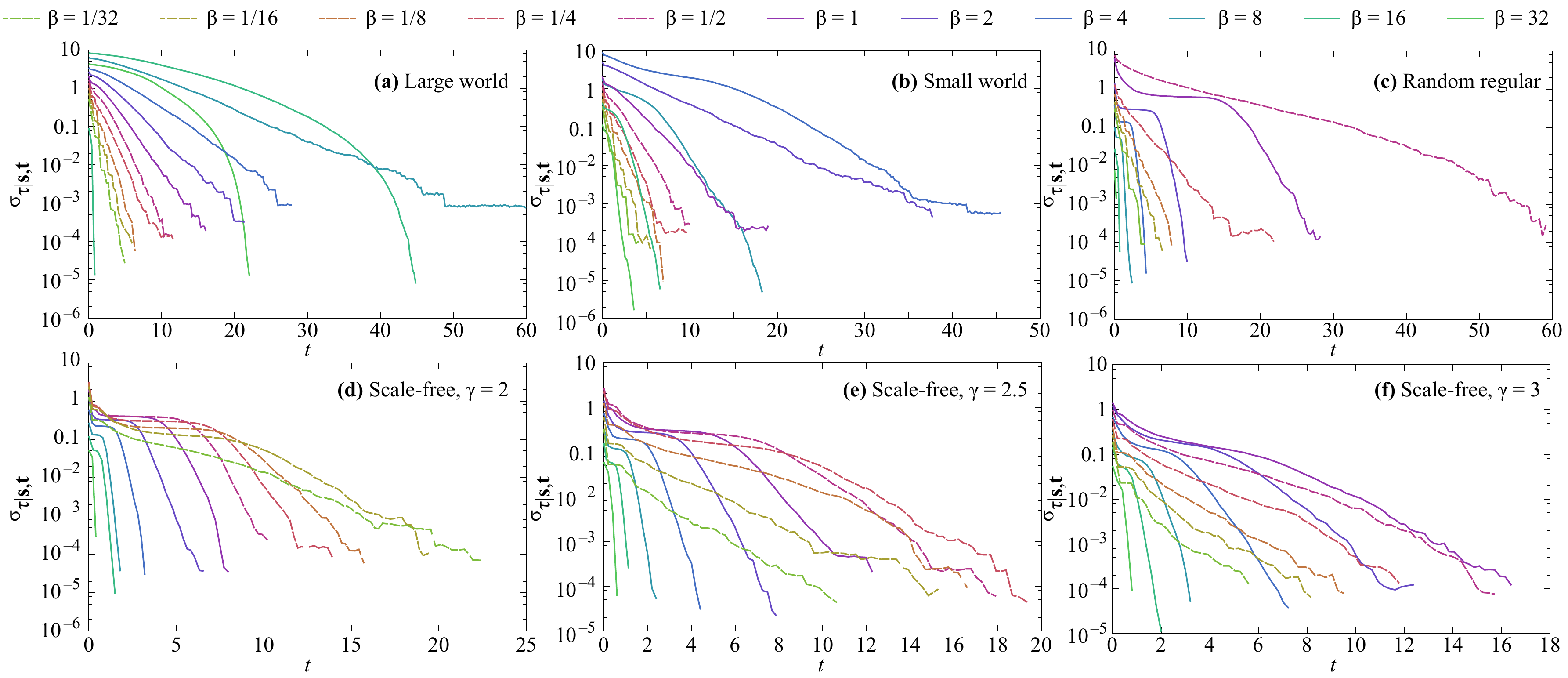}
\caption{(Color online) This figure corresponds to Fig.~\ref{fig:size}, but shows the standard deviation of the extinction time rather than the outbreak size.
}
\label{fig:time}
\end{figure*}

\subsubsection{Predicting the extinction time}

In addition to studying the (un)predictability with respect to outbreak size, we also study $\sigma_\tau$---the corresponding quantity for the extinction time $\tau$---in Fig.~\ref{fig:time}. The general picture from Fig.~\ref{fig:size} holds---the decay is roughly exponential. However, there is more structure in these curves. The decay fits the exponential form worse than what $\sigma_\Omega$ does. Several of the $\sigma_\tau$ curves for the random-regular and scale-free graphs plateaus at intermediate $\beta$ values. This means there are times when predicting the outbreak size gets more precise with time, but predicting how long the epidemics will last does not.

\section{Predictability given the sizes of the compartments}\label{sec:pred_comp}

In Sec.~\ref{sec:pred_state}, we studied the scenario in which we have maximum information about an outbreak. This is of course an idealized situation, and in this section, we turn to the more realistic scenario in which we know the number of people infected and recovered, but not who is in what state.

To assess the predictability based on the summarized information about the process, we redefine the measure of predictability $\sigma_{\Omega}$ by the standard deviation of $\Omega$ conditioned on the number of I vertices ($I$), R vertices ($R$), and I or R vertices.
We denote the measures redefined as $\sigma_{\Omega | I}$, $\sigma_{\Omega | R}$, and $\sigma_{\Omega | I+R}$, respectively.
In other words, $\sigma_{\Omega | I}$, for example, quantifies the uncertainty of the final outbreak size $\Omega$ conditioned on the fact that we know the $I$ value at time $t$. If $\sigma_{\Omega | I}$ of $t$ takes a small value, the final outcome of the epidemic process is predictable at time $t$ in the sense that the fluctuation in $\Omega$ conditioned on $I$ at $t$ is small.
Note that, because the total number of vertices is fixed to $N$, $\sigma_{\Omega | I+R}$ is equivalent to that conditioned on the number of S vertices.
We calculate these measures as follows (we take $\sigma_{\Omega | I}$ as an example).
Instead of setting a master sequence as we did in Sec.~\ref{sec:pred_state}, we first  simulate $10{,}000$ outbreaks with the initial seed chosen randomly on the given network realization.
We pick out the outbreaks with the same $I$ value at $t$ among the outbreaks and calculate the standard deviation of $\Omega$.
We do this calculation for all $I$ values observed at $t$.
Finally, we obtain $\sigma_{\Omega | I}$ of $t$ by calculating the expectation value of  the standard deviation of $\Omega$ over the distribution of the $I$ value observed at $t$.
For deriving $\sigma_{\Omega | R}$ and $\sigma_{\Omega | I+R}$, we perform the same calculation based on $R$ and $I + R$.

\begin{figure*}
\includegraphics[width=\textwidth]{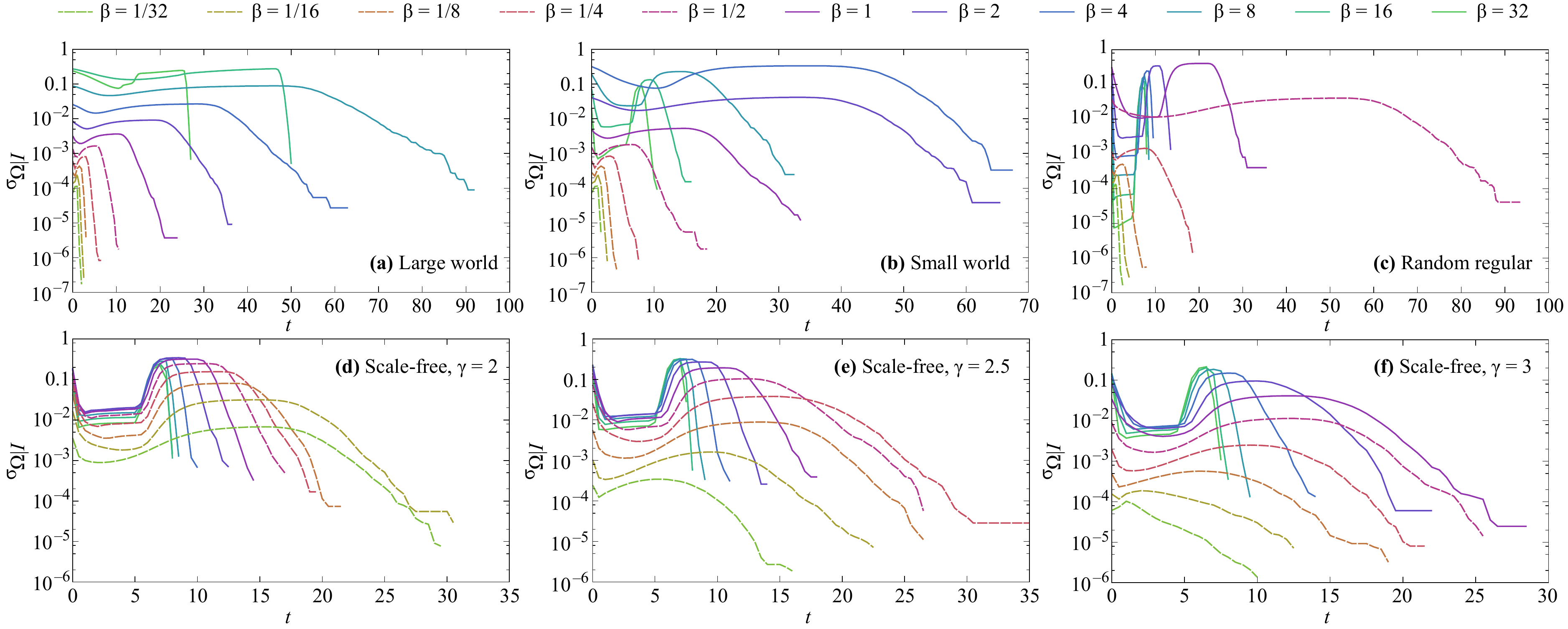}
\caption{(Color online) Standard deviation of the final outbreak sizes conditioned on $I$, that is, the number of infected vertices at $t$. Otherwise (the parameter values, number of averages, etc.)\ the plot corresponds to Fig.~\ref{fig:size}.}
\label{fig:size_I}
\end{figure*}

\begin{figure*}
\includegraphics[width=\textwidth]{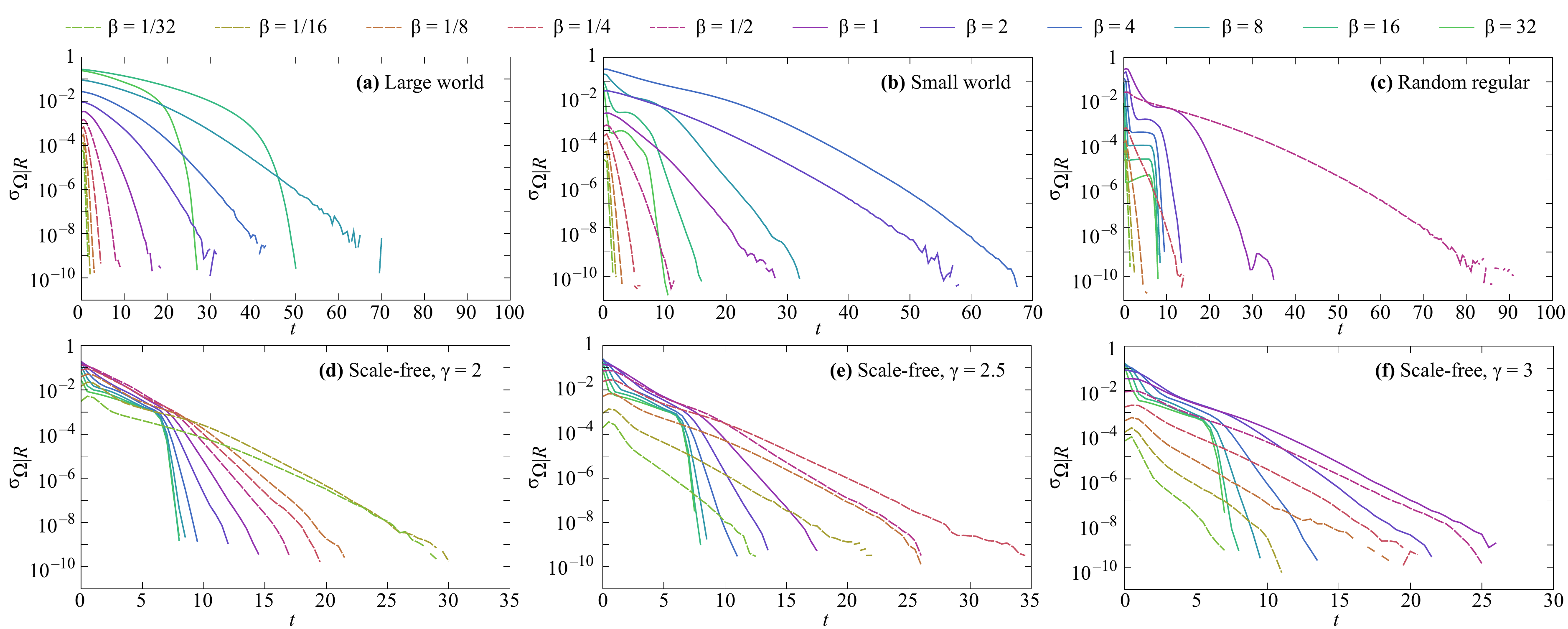}
\caption{(Color online) Plot corresponding to Fig.~\ref{fig:size_I} for the standard deviation of the final outbreak sizes conditioned on $R$, that is, the number of recovered vertices at $t$.}
\label{fig:size_R}
\end{figure*}

\begin{figure*}
\includegraphics[width=\textwidth]{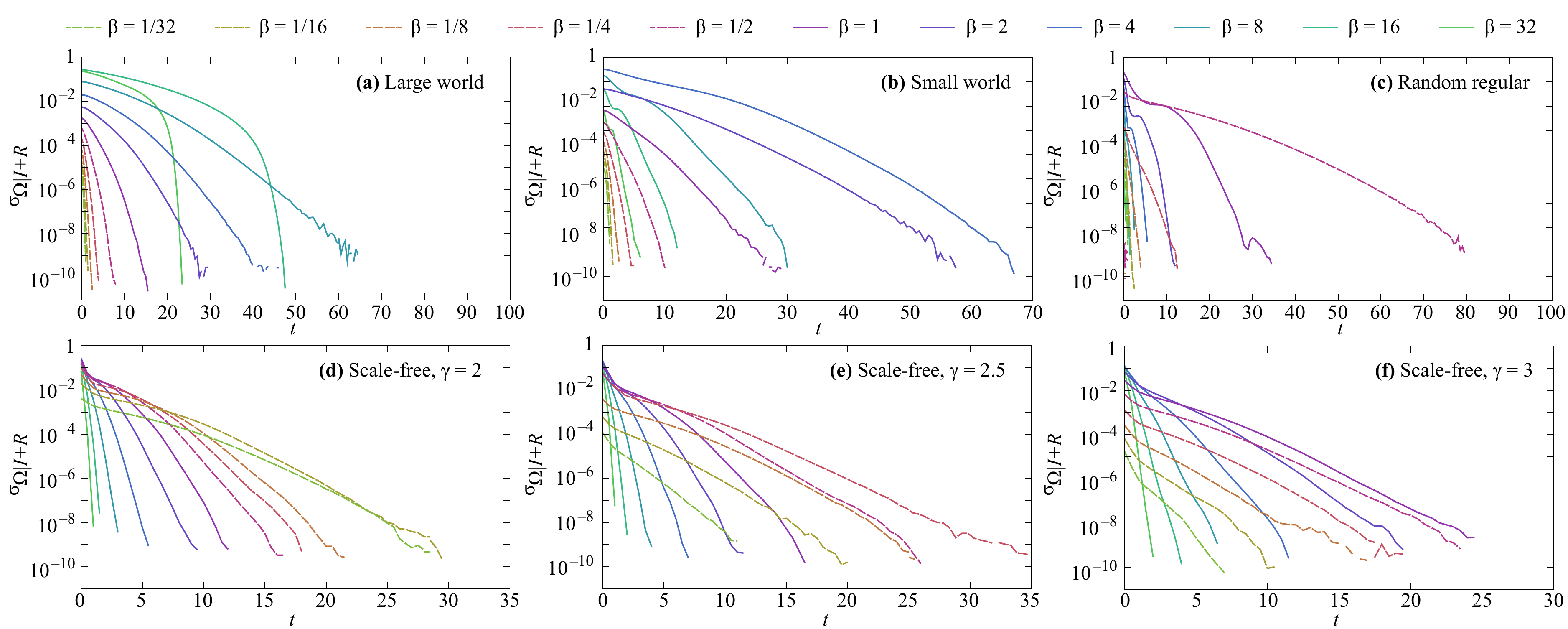}
\caption{(Color online) Plot corresponding to Fig.~\ref{fig:size_I} for the standard deviation of the final outbreak sizes conditioned on $I + R$, that is, the number of infected and recovered vertices at $t$.}
\label{fig:size_IR}
\end{figure*}

The resultant predictability measures $\sigma_{\Omega | I}$, $\sigma_{\Omega | R}$, and $\sigma_{\Omega | I+R}$ are plotted as a function of time~$t$ in Figs.~\ref{fig:size_I}, \ref{fig:size_R}, and \ref{fig:size_IR}, respectively.
Note that we use the  same set of network realizations that we used in Sec.~\ref{sec:pred_state}.
It also should be noted that the ranges of extinction time shown in Figs.~\ref{fig:size_I}, \ref{fig:size_R}, and \ref{fig:size_IR} are larger than those shown in Figs.~\ref{fig:size} and \ref{fig:time}, simply because we consider a larger number of simulation outbreaks than we did in Sec.~\ref{sec:pred_state}.
To be precise, here we consider $10^7$ simulation runs, i.e., $10^4$ runs on each of $10^3$ network realizations.

For all of the network models, we observe a bump in $\sigma_{\Omega | I}$ (see Fig.~\ref{fig:size_I}).
When we increase the $\beta$ value, we find the bump is located at the range of larger $t$ up to a certain $\beta$ value and is located closer to $t=0$ when we further increase $\beta$.
In particular, for scale-free networks with $\gamma = 2$ (Fig.~\ref{fig:size_I}(d)), we observe that the bump is located at the largest $t$ when $\beta = 1/32$.
The presence of a bump in the $\sigma_{\Omega | I}$ curves can be explained as follows.
At the beginning of the process, $\sigma_{\Omega | I}$ decreases with time~$t$ because the standard deviation of $\Omega$ among the outbreaks with the same $I$ values gets smaller as the outbreaks are separated into ones rapidly dying out and those eventually spreading to a large scale.
In the middle term of the process, $\sigma_{\Omega | I}$ increases with time~$t$ because a number of outbreaks, regardless of $\Omega$, fall into the range of the small $I$ values near extinction.
In the end of the process, $\sigma_{\Omega | I}$ decreases with time~$t$ because all the outbreaks die out and become perfectly predictable (as $I=0$).
These observations suggest that $\sigma_{\Omega | I}$ is a good predictor early in the outbreak, but that it becomes less reliable with time.

\begin{figure}
\includegraphics[width=\columnwidth]{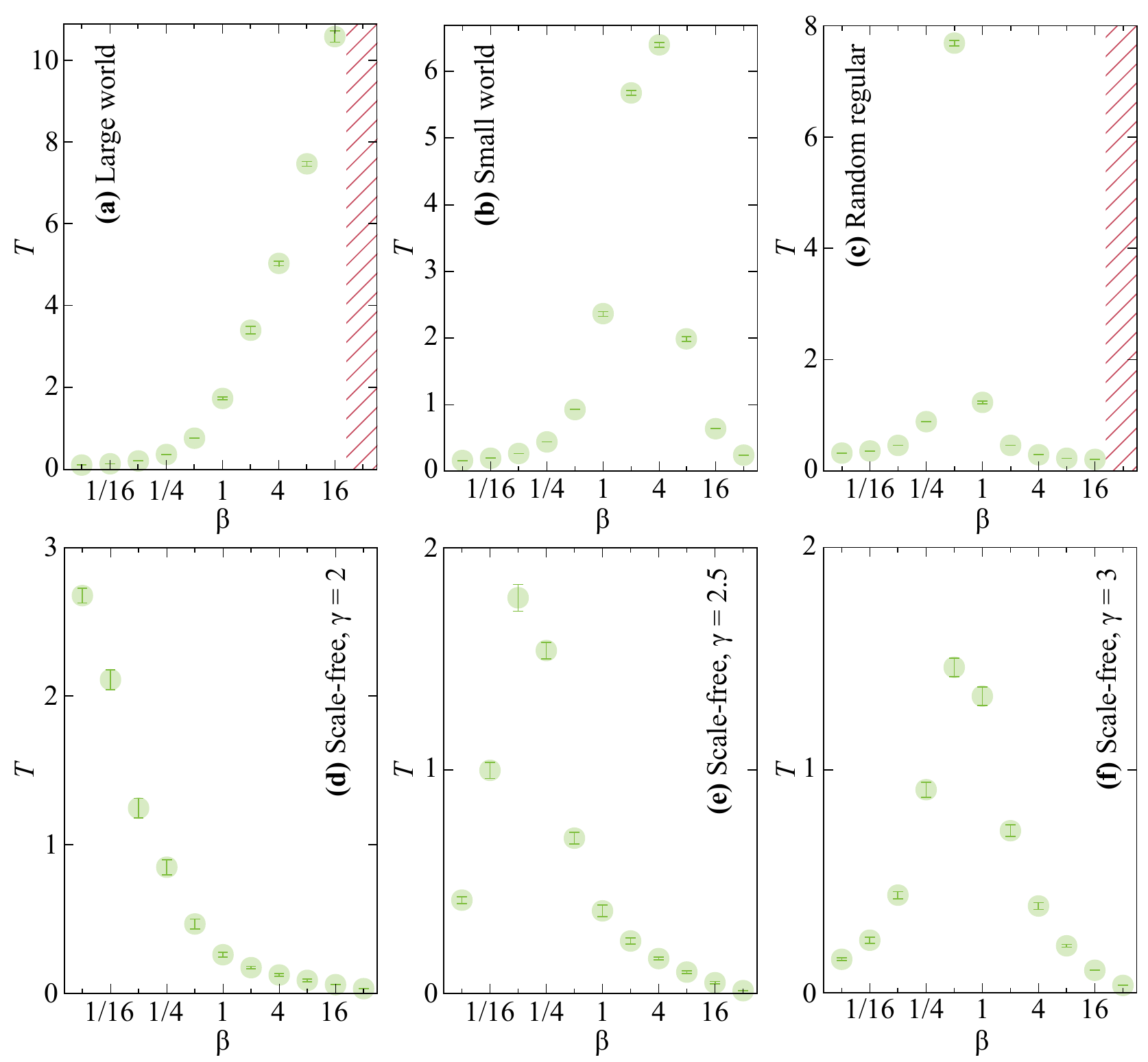}
\caption{(Color online) The decay constants from nonlinear least-squares fits to exponential forms of the curves of $\sigma_{\Omega|I+R}$ shown in Fig.~\ref{fig:size_IR}. The shaded regions indicate where $\sigma_\Omega$ does not fit an exponential function. The error bars represent standard errors.}
\label{fig:exp_fit}
\end{figure}

By contrast to the nonmonotonical behavior of $\sigma_{\Omega | I}$, the other measures $\sigma_{\Omega | R}$ and $\sigma_{\Omega | I+R}$ are nonincreasing functions with time except for the range of very small $t$ (see Figs.~\ref{fig:size_R} and \ref{fig:size_IR}).
This difference between $\sigma_{\Omega | I}$ and $\sigma_{\Omega | R}$ ($\sigma_{\Omega | I+R}$) can be explained with the intrinsic behavior of $I$ and $R$ as follows.
It is well known that, during the epidemic process, the $I$ value increases and then decreases to zero over time \cite{nwkepi_pastor} (also see Fig.~\ref{fig:size}).
This nonmonotonicity of $I$ results in the bump of the $\sigma_{\Omega | I}$ curves, as we discussed above.
On the other hand, the $R$ ($I+R=S$) value increases (decreases) monotonically over time \cite{nwkepi_pastor}.
If we observe a large $R$ value at $t$, we expect a larger $\Omega$ value in the end of the process.
In this way, this monotonicity of $R$ and $I+R$ results in the monotonic decrease of the $\sigma_{\Omega | R}$ and $\sigma_{\Omega | I+R}$ curves.
The results shown in Figs.~\ref{fig:size_R} and \ref{fig:size_IR} imply that the predictability of the SIR processes increases with time if we know the $R$ value (and the $I$ value in addition).
In particular, the $\sigma_{\Omega | I+R}$ curves exhibit the dependence on different $\beta$ values, which is very similar to the $\sigma_{\Omega}$ curves shown in Fig.~\ref{fig:size}.
This observation is supported by the decay exponents of the $\sigma_{\Omega | I+R}$ curves fitted with an exponential form (Fig.~\ref{fig:exp_fit}), which indicate a dependence on $\beta$ similar to that of $\sigma_{\Omega}$ (see Fig.~\ref{fig:t}).
In other words, with such a summarized value about the vertices' states, we can obtain the same accuracy in our predictions (in terms of $\sigma_{\Omega | I+R}$) as  when we have the perfect information about the system, for the network models we consider.

\section{Summary and discussion}
In this paper, we numerically investigated the predictability of outbreaks of the SIR epidemic process on static network models. We used the  standard deviation of the final outbreak sizes conditioned on the state of vertices at a given time as a key-quantity for predictability (technically speaking unpredictability, since zero standard deviation means that the outcome is perfectly predictable).
We considered the two scenarios of the information available. First, that we have perfect information of the system (where we know the exact state of the vertices $(\mathbf{s},\mathbf{t}^\mathrm{I})$). Second, that we have partial  information---to be precise, only the number of infected $I$ or recovered $R$. In the first scenario, the standard deviation $\sigma_{\Omega | \mathbf{s},\mathbf{t}}$ decays like an exponential function of time, or slower, for all the types of network models we considered. 
This result is consistent with the previous theoretical result that the characteristic quantities of the SIR process asymptotically converges to deterministic functions~\cite{janson_etal}.
The time constant of the exponential decay, however, is highly model-dependent.
We saw a monotonic increase of the decay exponent for the large- and small-world networks,
whereas a monotonic decrease was observed for the scale-free networks with $\gamma=2$.
For the random regular graph and the scale-free networks with $\gamma=2.5$ and $3$,
there is a single peak of the decay exponent at an intermediate value of $\beta$.
These results shed new light on the notion of predictability of the SIR process, i.e., the uncertainty in epidemic outbreaks is a nontrivial question even if we possess the information of contact network structure and vertices' states.

For the second scenario, the standard deviations $\sigma_{\Omega | R}$ and $\sigma_{\Omega | I+R}$ decay with exponential tails, while $\sigma_{\Omega |  I}$ exhibits a nonmonotonic change in time. Specifically, for large transmission rates it has a peak for intermediate times. In other words, some time after the outbreak starts, we can expect that knowing e.g.\ the prevalence becomes less valuable. It is also in this region where there is a large difference between the two scenarios, i.e., knowing the exact configuration of infectious, susceptible, and recovered individuals improves the predictability a great deal. Another way of seeing this is that for most parts of the parameter space, the additional information in our first scenario is rarely of much use, although it is much more information ($2N$ numbers as opposed to one number).

Connecting back to our starting point---how could real outbreaks be hard to predict when the SIR model itself converges to deterministic quantities---we see, as mentioned, that for some parameter values, assuming only aggregate information, the convergence is slow. A possible answer is that the discrepancy comes from that in practice we only have this kind of aggregate information. On the other hand, for many parameter values, also in the case of population-level information, the decay of predictability is very fast (also compared to the duration of the outbreak). For this reason, we also believe that the SIR model (i.e., the assumptions behind it), to some extent, underestimates the outbreak diversity.

This work paves the way for more detailed investigations of predictability of epidemic processes. For example, our work assumes that we know the starting point of the epidemic outbreak, which is fairly unrealistic. It would be interesting to investigate a situation in whcih that assumption is relaxed. This would connect our line of research to the question of identifying the source of an outbreak~\cite{source0,source1,source2} or reconstructing likely transmission trees~\cite{tree1,tree2}.

\begin{acknowledgments}
P.H. was supported by the Basic Science Research Program through the National Research Foundation of Korea (NRF) funded by the Ministry of Education (2013R1A1A2011947).
We thank Takehisa Hasegawa for comments.
\end{acknowledgments}

\end{document}